\documentstyle[epsfig,12pt]{article}
\begin{document}

% title

\begin{center}

{\LARGE\bf Electron Beam for LHC \\
 }
    
\vspace*{6mm}
\vspace*{6mm}

\vspace*{6mm}
\vspace*{6mm}
\vspace*{6mm}
% authors

{\it Mieczyslaw Witold Krasny,\\
     LPNHE, University Pierre et Marie Curie, Paris, France \\ }

\vspace*{6mm}
\vspace*{6mm}
\vspace*{6mm}
\end{center}

%/abstract
{\it Abstract: A  method of delivering a 
monochromatic electron beam to the LHC interaction
points is proposed. In this method,  heavy ions are used as carriers of projectile
electrons. Acceleration, storage and  collision-stability
aspects of  such a hybrid beam is discussed and a 
new beam-cooling method is presented. 
This discussion  is followed by a proposal  of 
the  Parasitic Ion-Electron collider at LHC (PIE@LHC). 
The PIE@LHC provides an opportunity, for the present LHC detectors,  to   
enlarge the scope of their research program  by 
including the program of electron-proton and electron-nucleus
collisions with minor  machine and detector investments.}

\section{Introduction}

The LHC detectors will explore the 
high energy frontier of proton-proton, proton-nucleus
and nucleus-nucleus collisions.
From the perspective of 
the Standard Model the r\^ole of proton and nuclear beams 
is confined  to   
that  of carriers of  virtual but  Lorenz-frozen,
Wide-Band-Beams (WBBs) of 
the basic,  point-like building blocks 
of matter,  and of their interaction quanta.  

Hadronic colliders are optimal 
for generic exploration  of interactions of a large 
variety of the Standard Model particles, over a 
large momentum range, which  extends up to the
momentum of the carriers of partonic beams. Their merits are 
complementary to those of  electron-positron colliders,
which employ  better-controlled,  but  less-luminous Narrow-Band-Beams (NBBs) 
of the Standard Model particles, colliding  in a cleaner 
environment.

The electron-proton (electron-nucleus) colliders can neither  
compete with the hadronic colliders in exploring the small-distance  
frontier of interactions of the basic constituents of matter,  
nor with the electron-positron colliders in the measurement precision.
Nevertheless, they are very useful  for a precise mapping of the emittance and  
momentum distributions of partonic WBBs generated by colliding hadrons.
Such a mapping will remain indispensable as long as  
the quantitative predictive power of the 
Standard Model will be confined  to the perturbative calculation
methods. 
It must involve unfolding of the effects
of space-time confinement of virtual bunches of partonic beams, 
and the effects due to strongly-interacting medium in which they are colliding.

The experimental programme of 
the LHC collider could be significantly
enriched, and the measurement precision could be increased,  
by adding  a monochromatic electron beam  to the LHC beam inventory,  
in particular,  if such a beam could be delivered   
to the Interaction Points (IPs) of the LHC detectors,
for collision with the LHC hadronic beams. 
If 
provided, it could  
allow the LHC measurements to be largely 
independent of the 
partonic-distribution-function (PDF) extrapolation-technology.
This technology assumes the validity
of the factorization (WBB-invariance) of the Standard Model
hard processes over the full kinematic domain, including regions
where it has never been tested.
In addition, it relies  upon the   
phenomenological modeling of the energy and of the 
nuclear dependence of partonic distribution and upon
the energy-independent partonic-beam emittance.

If the  electron-proton and the electron-ion
collisions could be observed and recorded by the LHC detectors
concurrently with  hadronic 
ones, they would  
provide, for the latter, a  high-precision control of the time dependent 
detector calibration.   
Their studies could  allow for a  precise,
detector-specific  unfolding   
of the non-perturbative QCD effects,
which are present in  the relation
between the fast-moving coloured constituents of matter and 
the observed jets.
A monochromatic electron beam accompanying partonic WBB in hadronic  
colliders could thus play a similar  r\^ole to that of  
a monochromatic electron beam 
which produces  the partonic NBB in the electron-positron colliders.

The merits of the electron beam 
must be weighted against an extra cost and  against 
a possible interference of the corresponding                 
electron-proton  (electron-nucleus) 
collision programme 
with the standard LHC programme. In this paper, I shall assume that 
the electron beam will not be accepted unless  it is delivered 
with no extra cost.
In addition,  I shall require  
the bunch structure of the electron beam to  be identical to that of
the standard LHC beam such that the LHC detectors could register the
electron-proton (electron-ion) interactions without any 
hardware modification,  in particular with no readjustments 
of the detector electronics. 
Moreover, storage of the electron beam must not degrade 
the quality of the standard LHC beams, and its presence  must not  
be in conflict  with the foreseen operation mode of the LHC collider.  
The electron beam  must be, thus, fully parasitic. 
Last, but not least,  I shall require the energy of the 
electron beam to be optimal for a high precision mapping 
of partonic WBBs at the LHC,  
in the most difficult, small $x_{Bj}$ region, 
using the unmodified  LHC 
detectors.
%which are  optimal for those of partonic interaction in which
%`the collision Center-Of-Mass reference system is approximately at rest
%in the detector refence frame.       

The goal of this paper is to propose a method of delivering 
a monochromatic electron beam to the IPs
of the LHC experiments satisfying 
all the above requirements. The present paper summarizes the preliminary 
feasibility studies of the proposed  method. The 
technical details will  be presented elsewhere.

\section{The idea}

The basic idea is very simple. It consists of using  
heavy ions  as the carriers 
of the monochromatic beam of electrons. 
Electrons are attached to the atomic nucleus  by means of Coulomb 
attraction and they form a partially ionized atom.
The distance between the  electrons and  the atomic nucleus,
even in the case of lowest energy states of the highest-Z atoms,
is significantly  larger than the range 
of strong interactions. This  allows the electrons 
and the nucleus to be considered
as free particles 
in large momentum transfer collisions.
The electron collisions, at the energies of LHC collider, are  thus 
{\it almost} unperturbed
by the presence of the nucleus.\footnote{The 
electromagnetic radiative corrections
due to the soft photon emission have to be calculated assuming 
an effective cut-off reflecting the quantum structure  
of atomic levels.}     

An atomic nucleus  carrying $k$ electrons can be stored at 
the energy which is limited by the maximum  
magnetic field of the bending magnets:
\begin{eqnarray}
E_{ion}= E_Q \times (Z-k).
\end{eqnarray} 
The $E_Q$ represents 
the energy-per-charge machine parameter, related to the 
strength of the  magnetic field,  and $Z$ represents  the number of protons
in the nucleus. Each  electron carries  the 
longitudinal momentum of: 
\begin{eqnarray}
p_e = M_e/M_{ion} \times E_Q \times (Z-k),
\end{eqnarray}
where $M_e$ and $M_{ion}$ are, respectively, the electron and 
the ion masses.
Note,  that smearing of both the longitudinal 
and the transverse momentum of the electron,
due to its orbital motion, contrary to the Fermi 
motion of the nucleons within a  nucleus,  can be neglected 
for the LHC collider energies. The
electron beam  can be considered, thus,   
with a high precision as monochromatic.

If, for example, two electrons would be  transported to the 
LHC IPs  by the lead ions to  
collide with the LHC protons at the  top energy, then  the maximum CM-energy of 
the electron-proton collisions would be $ \sqrt{s} \approx  200$ GeV.
This value is close to the energy of
the HERA collider. If 
such a collision-configuration would be  realized  
at the RHIC collider at the BNL,  the CM-energy 
of the electron-proton  collisions would be smaller  than 
those   of the SLAC fixed target 
deep-inelastic experiments, where the partonic
picture of nucleons was established.

As far as the achievable electron-proton collision energy is considered,
the LHC is, thus,  the first hadronic collider,  in which interactions
of electrons carried by the ions could  be efficiently used to 
map, over  a large kinematic domain,  partonic structure of nucleons and nuclei. 
A large asymmetry of the energies of the electron beam,  $ E_e \approx 1.5$ GeV,
with respect to the  
proton beam, $E_p =  7 000$ GeV,  makes such a collision-scheme particularly well suited
for mapping of small $x_{Bj}$ region using  the existing 
LHC detectors. As an example,  for collisions of massless partons, 
carrying a fraction 
of $2 \times 10^{-4} $ of the nucleon momentum,   with such an electron beam,  
the electron-parton collision CM-frame is
at rest with respect to the laboratory  frame of 
the LHC detectors. This configuration allows for an optimal 
measurement of deep inelastic scattering kinematics in the
precise barrel sections  of the existing detectors. 

There are two ways of delivering partially ionized atoms to the LHC ring.
The first one is to use an electron capture foil in the transfer line 
between the SPS accelerator and the LHC storage rings.
This method is not efficient and will not be 
discussed in this paper. The second, advocated method consists of modifying the 
standard stripping sequence of the LHC ion beams. 
For example, stripping from the the $Pb^{54+}$
charge state  to the  $Pb^{82+}$ charge  state (fully-stripped ion), 
foreseen at the transfer line between the PS and the SPS,
would have to be replaced by a stripping procedure which leaves
 $k$ electrons. Such a  beam would have to 
be transfered to the SPS and, subsequently, to the LHC and accelerated to the 
top energy.
The optimal value of $k$, discussed in 
more details in the following sections,
should represent an  optimal trade-off between  
the acceleration and the 
collision stability of the beam at the  largest 
possible energy - requiring $k$ to be
be small,  and the largest possible 
equivalent current of the electron beam
- requiring  $k$ to be large.
In addition,   the stripping loses due to the acceptance
of the only one charge state at each stripping stage,  
have to be minimized  by limiting the number of stripping stages 
and by stripping on a shell-by-shell basis.

The idea of using ions as carriers of a monochromatic beam of electrons  
seems to be  rather straightforward. What is far from being straightforward
is how to form, accelerate, store 
and collide such hybrid beams at the top LHC energy.
 
The beams of partially-stripped ions are particularly fragile
and have to be treated with  a special care. 
Exposed to the strong magnetic field of 
the LHC bending magnets,  they have to survive 
a static  electric field,  which is
20 times stronger  than that binding an electron on the K-shell
of the hydrogen atom. 
When accelerated to the top LHC energy, 
the temperature  
of bunches of partially-stripped ions
must be stabilized, 
such that the ions  are not exposed to temperatures 
exceeding  their  characteristic ionization temperatures.
To be stored over a long time at the top energy,  they need 
to be strongly bound
to minimize losses due to interactions with 
the residual gas in the machine. The luminosity of colliding 
beams of partially-stripped ions should be sufficiently high to allow for
a satisfactory 
statistical precision of mapping of the WBBs at the LHC,  
but not too high for an excessive beam-collision stripping rate. 
The last two constraints must be give a special attention in the 
case of superconducting storage rings.   
Even if one electron is lost,  the magnetic rigidity of the carrier 
ion is changed and the ion hits  the beam pipe 
within a fraction of a microsecond.  If the rate of
the electron losses is high, then  such desynchronizing 
of their parent ions may cause quenching  of the magnets.

At first sight, dealing with such beams in 
superconducting storage rings seems to be 
hopeless. This is presumably the reason why it has never been considered.
All the above constraints 
impose, indeed, very stringent requirements on 
what beams can be formed, and how they must be
accelerated, stored,  and collided.
These requirements are discussed in the following sections.

\section{Survival of partially-stripped ions in the LHC  lattice}

Beam particles circulating in the LHC storage rings are exposed 
to a strong  magnetic field of superconducting dipoles.
The bending field $B$, 
when viewed in the rest frame system of stored
bunches,  is equivalent to an electric field $E$ with the strength 
of:
\begin{eqnarray} 
E = \gamma \times \beta \times c \times B,
\end{eqnarray}
and of the direction 
perpendicular to that of the magnetic field. 
The dipole 
field of 8.4 Tesla 
is equivalent, for ultra-relativistic beam particles 
moving with velocity corresponding 
to the Lorenz-$\gamma$ of  3000,  to the   
electric field of a strength: $E~=~7.3 \times 10^{10}$ V/cm.
The field of such strength will not only modify the atomic 
energy levels of partially-stripped ions (Stark effect)
but, may, in addition,   strip the electrons.

Rausch and Traubenberg \cite{Rausch},  while investigating  the effect of the electric field
on the atomic Balmer lines,  observed that each spectral line ceased to exist
above a certain field strength.
The suppression of spectral lines was the first observation of ionization
of atoms by strong electric fields.

The quantum-mechanical description of  
the mechanism of quenching of spectral lines 
has been presented 
in the work of Lanczos \cite{Lanczos} and Oppenheimer \cite{Openheimer}.
Spectral lines are suppressed if the probability of ionization
of the initial excited state of an atom is greater than the probability
for its radiative de-excitation.
Since the atomic de-excitation time is typically of 
the order of $10^{-8}$ seconds, the quantum-mechanical tunneling 
probability of an atom to lose  
over an atomic unit time ($ 2.4 \times 10^{-17}$ seconds) one electron 
must be smaller than  $10^{-9} $.
This limit determines, for each atomic energy level, the maximum 
field strength,  at which an atom in a given charge state can survive
or,  conversely, for a given field  strength, which charge states
of partially-stripped ions 
can be stored at a given energy.

The results of the calculations, which are relevant 
for the present paper,  
are  summarized below.
Light ions, up to oxygen, can be stored in the LHC ring
only in the fully-stripped charge state.
High-$Z$ partially ionized atoms  can survive the LHC magnetic field
in a wide range  of charge states, provided that
the binding energy of the least-bound electron
is kept above $E_b \approx 0.9$ keV. 

In the following, I  shall restrict the
discussion  to the case of lead ions
as potential candidates of carriers of the electron beam. 
The main reason is that the fully-stripped lead-ion beam is already foreseen 
in the LHC experimental programme.\footnote{All the 
arguments and conclusions presented in this paper hold if lead ions
are replaced by gold ions (the BNL heavy-ion beam).}
I  shall also restrict 
the discussion to the charge state $Pb^{80+}$
which maximizes both the electron beam energy 
and intensity.
In such a charge state, 
two electrons remain attached 
to the $Pb$ nucleus. 
These electrons can be safely transported  
not only on  their K-shell orbit, but 
also in a wide range of excited states, provided 
that the energies  of their excitations will not exceed 
the critical one. This  critical 
energy for $Pb^{80+}$ turns out  to be  
equivalent to the binding energy of  
the excited $n_{max} =  10$ state of the  $Z=82$,  single-electron
Rydberg atom.

The $Pb^{80+}$ ions can be thus considered as robust
candidates for the carriers of electrons resistant to 
various environmental perturbations,  
which may  result in their excitations.

\section{Survival of bunches of partially-stripped ions}

\subsection{Temperature}

Ions are accelerated and stored in bunches. 
The evolution of bunch sizes is  driven by 
the RF-system and by the 
beam focusing optics. 
The motion of the beam particles within their bunches 
can be characterized by the bunch temperature.\footnote {Expressing the 
beam particle motion in terms
of an equivalent temperature requires the average collision time between
the beam particles to be small with respect to the acceleration and the 
storage times. Note that, in this context, 
the frequency of the intra-beam collisions of partially-stripped
ions is significantly higher than that of the protons.}
In the following, I  shall concentrate on the 
transverse motion of the beam particles.
The longitudinal motion and its coupling to   
the transverse  motion  
is omitted in this paper for simplicity.
 
The transverse temperature of bunches can be expressed in terms
of: the normalized emittance of the beam, $\epsilon _{N}$, the average
(over the ring) transverse size of the bunches, $\sigma_{h,v}$,  the mass 
of the ions, $M_{ion}$,  and the ion velocity $\beta$ as: 
\begin{eqnarray}
kT_{h,v}= M_{ion} \times c^2 \times \beta ^2 \times (\epsilon _{N}/\sigma_{h,v}). 
\end{eqnarray} 
The critical issue  of accelerating bunches of partially-stripped ions becomes
apparent when the temperature of the LHC bunches
is calculated. 
The transverse temperature at the injection to the LHC reaches 
equivalent energy of 1 MeV and rises linearly 
with increasing $\gamma$ of the beam.
Even if initially carried on   
the K-shell of the highest-Z ions, the electrons are stripped  at such temperatures.

\begin{figure}[ht]
\begin{center}
\leavevmode
\epsfig{file=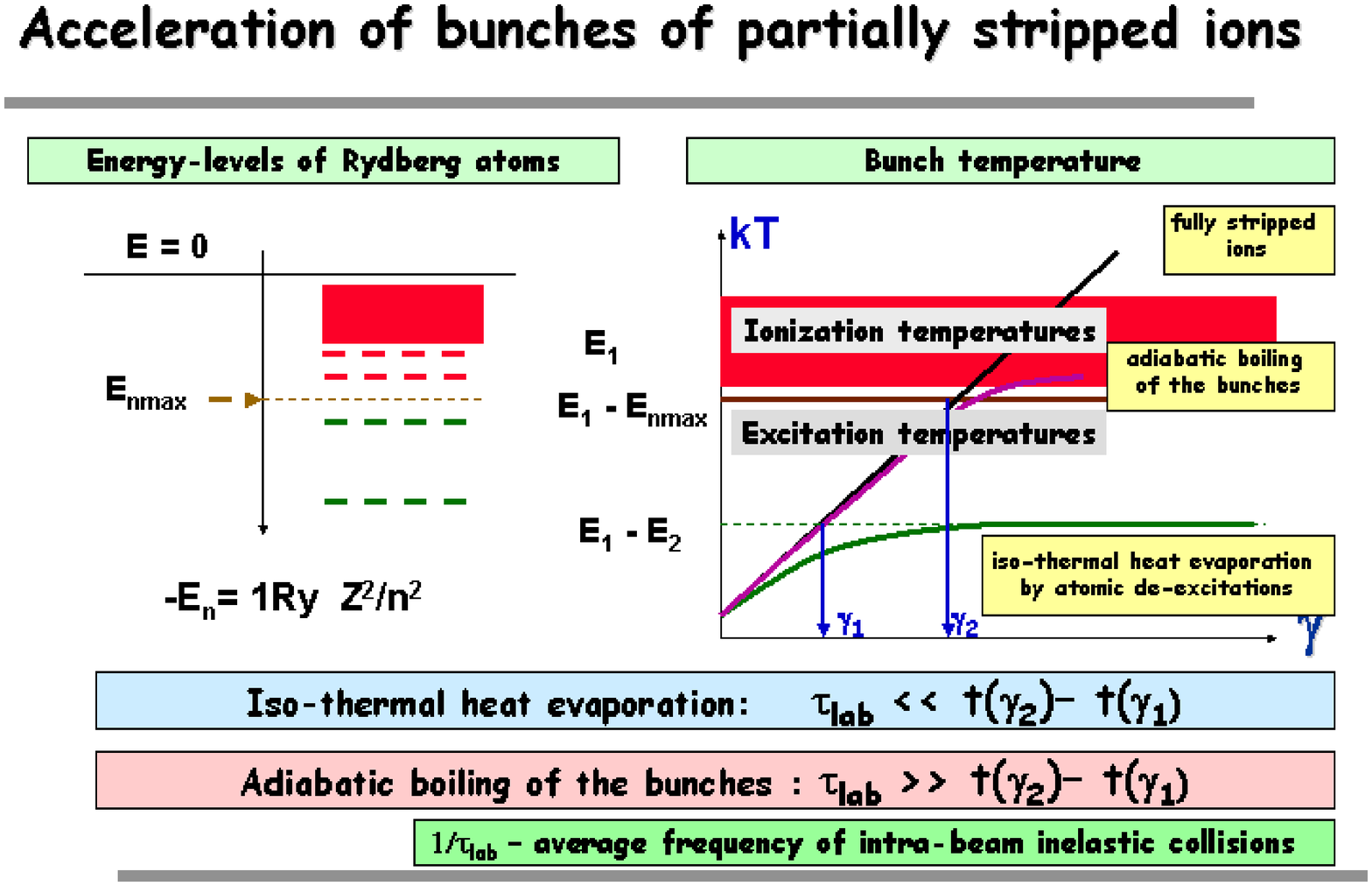,angle=0.,width=14.0cm}
%\vspace{1.0cm}
\end{center}
\caption{The temperature evolution of 
the accelerated bunches of partially-stripped ions.}
\label{acceleration}
\end{figure}

A method of stabilizing temperatures of bunches  below 
the ionization temperature must be invented for such a  beam. 
The method, which is  proposed here is based upon a concept of 
the controlled,  iso-thermal
heat evaporation by light emission.
This method is illustrated  in Fig. \ref{acceleration}
and explained in the following.

\subsection{Temperature evolution of bunches of fully and partially-stripped ions} 

The energy levels of an atom with the nucleus charge $Z$ 
can be subdivided into two subsets:
the energies below and above $E(n_{max})$. The $E(n_{max})$ is
the energy of the atomic level above which the ion
is  spontaneously ionized by the machine B-field.
If an electron carried by the lead ion 
on its K-shell ($n=1$ state) is excited by an  inelastic collision
to the $n>n_{max}=10$ state, it will be promptly
stripped off.
%\newpage
%width=10cm,bbllx=-29988 ,bblly=274 , bburx=1012 ,bbury=30780}
%bbllx=-29988 ,bblly=274 , bburx=1012 ,bbury=30780}

At high beam energies, 
the temperature of the bunches of {\it fully}-stripped ions increases
with the Lorenz $\gamma$ factor of the beam. This dependence 
is analogous to an increase of the temperature of an ideal  gas,
in the adiabatic change of the gas volume. 
%\newpage
%width=10cm,bbllx=-29988 ,bblly=274 , bburx=1012 ,bbury=30780}
%bbllx=-29988 ,bblly=274 , bburx=1012 ,bbury=30780}
\begin{figure}[ht]
\begin{center}
\leavevmode
\epsfig{file=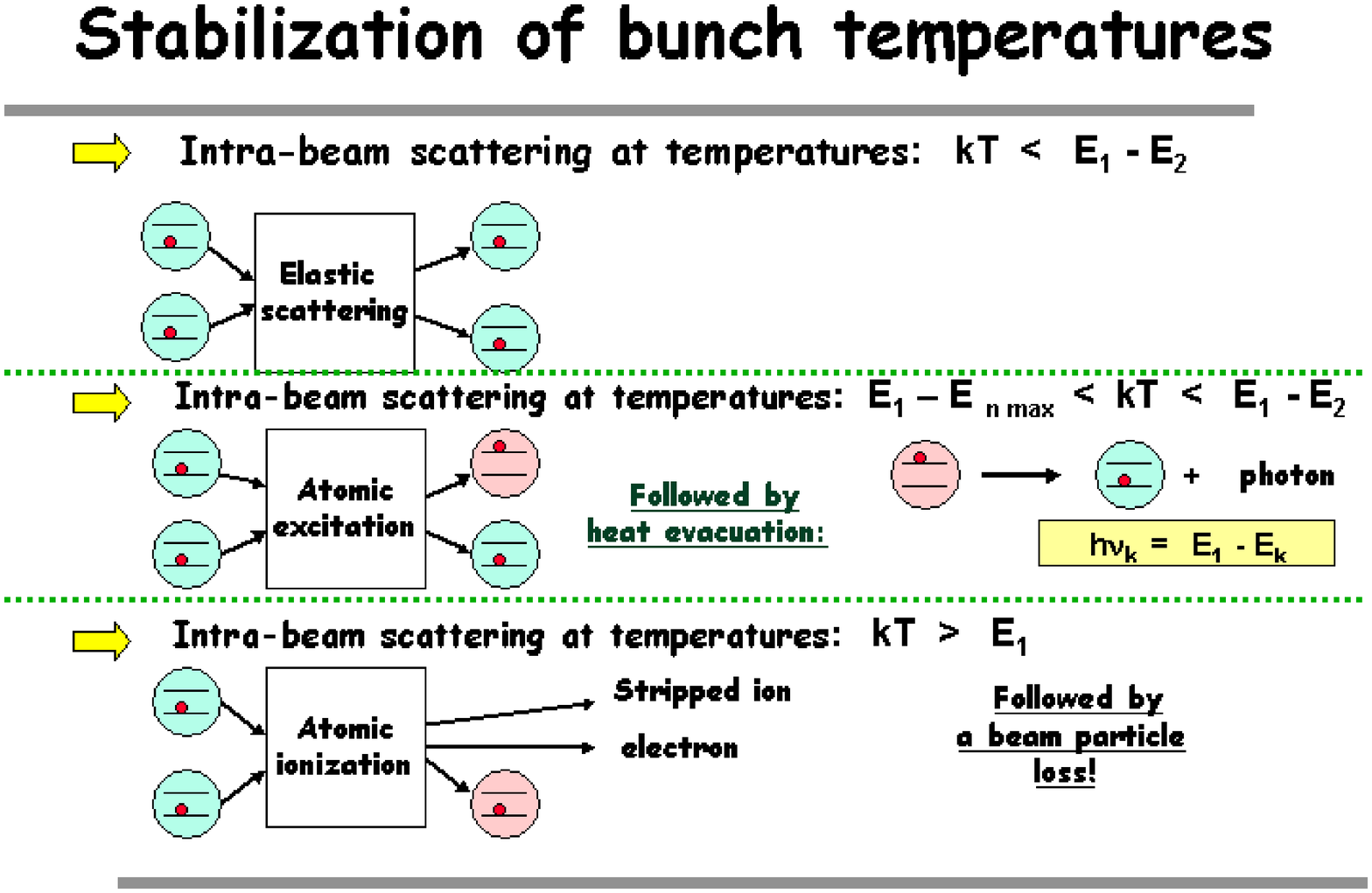,angle=0.,width=14.0cm}
%\vspace{1.0cm}
\end{center}
\caption{Intra-beam collisions of partially-stripped ions.}
\label{stabilization}
\end{figure}

The temperature evolution of the bunches  of {\it partially}-stripped 
ions will be affected by their  atomic energy-level structure in 
a way which is illustrated  in  Fig. \ref{stabilization}. 
At the early acceleration stage, when the bunch temperature is lower
than the difference between the ground state and the first excited
state of the ion, the beam collisions are solely  elastic. 
In this regime,
the temperature evolution of the beam of partially-stripped 
ions will be the same as that for  the fully-stripped ion beam.
At the acceleration stage,  when the bunch temperature becomes 
larger than the difference between the ground,  and the first 
excited state of an ion, but smaller than the difference between 
the ground state and the $n_{max}$ state, ions will collide
both elastically and inelastically. The inelastic collisions will lead
to atomic excitations followed by photon emissions.
Finally, if the temperature of bunches will be allowed to 
exceed the critical ionization level (the difference between 
the ground state and the $n_{max}$ state) 
 the inelastic collision of partially ionized atoms
may lead to their full ionization. 

\subsection{Fast acceleration of bunches of partially-stripped ions}
 
Given the above picture,  two methods  of accelerating bunches of
partially-stripped ions are  proposed. The first one consists 
of  fast acceleration of the beam. The acceleration gradient  
$d \gamma/dt$, in this method, is required to be  
sufficiently large  such that,  at any given moment of 
the beam acceleration, within the time interval between  
reaching the critical ionization temperature
and reaching the top beam energy, the bunches are not allowed to 
thermalize. Note, that the termalization time increases linearly
with the Lorenz $\gamma$ factor.
This  method employs an  effective ``dilution'' of the gas of
overheated ions due to slowing down the intra-beam
collisions as observed in the accelerator rest frame.
The resulting bunches will be,  however,  unstable - the 
inelastic intra-beam
collisions at the top energy will lead to stripping of electrons
and the beam will eventually be lost unless it is quickly 
cooled down at the LHC top energy.
The Doppler cooling \cite{Bessonov} 
may be used for this purpose.
Its efficiency would be significantly 
enhanced if the atomic
transitions rather than the nuclear excitation transitions are used.
It is interesting to note, 
that the LHC is the first collider where such a cooling 
can be done at the top beam energy using 
conventional, high power UV lasers.
Note, that  the $\gamma$ factor at the top LHC energy, driving 
the Doppler gain of the energy of the laser light, becomes  
sufficiently high to compensate 
the  $Z^2$ increase of the binding energy of atomic levels
of partially-stripped ions.

The fast acceleration method complemented by the 
laser cooling at the top LHC energy is indeed very attractive
but requires dedicated investments.
Such investments are unlikely be made at CERN solely   
for the purpose of cooling of partially-stripped ion beams. 
They may, however, be driven by the side gains
of such a cooling scheme (e.g., creation at CERN 
of a luminous, {\it monochromatic} and highly collimated X-ray beams in the 
MeV energy range).
    
In order to avoid adiabatic boiling of the bunches at the collision
energies, without external beam cooling, 
an alternative  acceleration method is proposed below.
In this method, the excessive heat of 
adiabatically compressed bunches   
is evacuated iso-thermally 
by means of radiative evaporation.
The acceleration rate of partially-stripped ions must be 
maximized at low energies to avoid the effects 
driving  the effective
increase of the bunch sizes. The ions must then be stripped 
as early as possible to the final beam-collision charge state.
The crucial point of the method is that, 
at the critical acceleration phase of the beam 
in its final beam-collision charge state,
when the bunch temperature reaches the atomic excitation
temperature 
($\gamma_{crit} = \gamma (T= (E_1 - E_2)/k)$),  the process 
of acceleration must be 
slowed down to allow for gradual heat evaporation.
The acceleration rate   
$d \gamma/dt$ should be 
reduced  to such a value
that at any moment of 
the beam acceleration process, spanning  
the time interval between   
reaching the critical ionization temperature
and reaching the top beam energy, the ions are 
kept at thermal equilibrium.
Under this condition,
inelastic collisions of ions could effectively convert the 
excessive kinetic energy of the most energetic
ions (occupying the tail of the Maxwell distribution)
into the energy of their excitation.
The energy absorbed by the atomic excitations 
will be subsequently  
evaporated
by the ion-de-excitation photons.
In the model of asymptotically-slow acceleration, 
with no other dissipative processes present,  the
asymptotic beam temperature should stabilize
at the temperature of $T_{max}= (E_1 - E_2)/k$.

\subsection{Numerical exercise}

The condition for the iso-thermal heat evaporation 
can be quantified by using an acceleration 
model specified in terms of the following parameters:

\begin{itemize}

\item
number of $Pb^{80+}$ ions  per bunch - $N_{ions/bunch} = 10^7$,
\item
normalized beam emittance - $\epsilon_{N} = 1.5 \times 10^{-6}$ m, 
\item
average $\beta$ over ring circumference - $<\beta > = 67$ m,
\item
bunch length in the storage-ring rest frame -$L_{bunch} = 7.5 $ cm, 
\item 
the atomic excitation cross section - $\sigma _{excit} = 10^ {-18}$  cm$^2$.
\end{itemize}
 
In such a model,  the temperature of lead bunches is expected to 
stabilize at the level of  68 keV, provided that 
following the fast acceleration phase, 
up to$\gamma \approx 10 $, 
the acceleration rate is reduced to the value of $d \gamma /dt < 0.8$/minute. 
In this model, slowing down of the acceleration rate must take place 
at the SPS. 
If the above model is applied to the acceleration of 
bunches of the gold ions at BNL,  the onset of the temperature
stabilization should be observed in AGS, 
where  bunches of $Au^{77+}$ 
carrying two unstripped electrons
are accelerated.
The processes of temperature stabilization
is expected to  reduce the normalized
emittance of the beam. Thus,  the emittance 
of the top-energy $Au^{77+}$ beam in AGS 
should be smaller than that of 
the beam fully-stripped $Au^{79+}$ ions. 
 
\subsection{Merits of the temperature stabilization}
 
If the temperature stabilization is confirmed experimentally, the low
temperature bunches of partially-stripped ions could  be accelerated up
to the top LHC energy with residual ionization losses due to stripping 
of the ions by the intra-beam scattering.
Moreover, the proposed  cooling technique  could be 
employed to improve the beam quality for 
the standard LHC collision programme
which uses fully-stripped ions. In order to  profit
from the temperature stabilization, the 
stripping of the two last electrons of the lead ions
would have to be postponed up to the beam transfer between the SPS and the LHC.  
The effective reduction of the beam emittance could allow for the reduction 
of the beam current at fixed luminosity. Equivalently 
it could allow  for increasing the $\beta^*$ value 
at fixed beam current and luminosity. But, first of all, 
it could allow to reduce the beam envelope providing cleaner beams.

\section{The lifetime of the beam of partially-stripped ions in the LHC collider}

\subsection{Ionization processes}

If thermally-stable,  partially-stripped 
ion bunches can be stored at the LHC,   
their lifetime will be governed  by the  
rate of collisions of ions  
with the residual gas in the ring.
The beam losses will be dominated
by the ionization processes in which 
an  electron is stripped-off from the 
carrier ion by a residual gas molecule.
In the collision phase, the electrons
carried by the ions will  also   
be stripped in the beam-beam collisions.
While the 
beam-gas losses can be
reduced by improving the machine vacuum, the 
beam-beam collision losses  represent an 
irreducible limit governed by the machine
luminosity. 

The K-shell ionization of ultra-relativistic ions has been discussed
extensively by R. Anholt and U.Becker \cite{Anholt}.
The ionization cross section can be expressed as the sum of the three 
contributions:  the Coulomb , the transverse, and  the spin-flip ionization.
The Coulomb cross section does not depend upon the collision
energy. The contribution of transverse ionization 
is most important in collisions with large impact parameters.
It increases logarithmically with increasing   $\gamma ^2$.
The spin flip contribution can be safely neglected in 
the following discussion.

The theory of Anholt and Becker has been successfully 
tested at $\gamma= 168$ by measuring $Pb^{81+}$
ionization cross section in gaseous Ar, Kr and Xe targets
\cite{Krause}. The predicted cross sections were found to be 
slightly larger that the measured values.
Thus, there is a safety margin in 
applying  this theory to calculate  the lifetime 
of the beam of partially-stripped ions at the LHC, by  extrapolating 
the existing measurements 
to different targets and higher collision energy.
Such  extrapolations  must take 
into account the difference between the ionization 
of the projectile particles by atoms (verified experimentally),
and by fully or partially-stripped ions. The ion charge screening 
corrections, which flatten the energy dependence
of the ionization cross-section, become small in the limit 
of $Z_c >> Z_t$, where $Z_c$ is the charge of the carrier 
of the electrons and $Z_t$ is the charge of the target.
In this limit K-shell-electron wave functions 
are  ``localized'' within a small transverse surface, with respect 
to the distances between the target electrons and target 
nucleus and projectile electrons ``see'' almost-unscreened target nucleus.

Another important aspect of the R. Anholt and U.Becker theory,
which will be employed  in the subsequent sections,  is the 
dependence of the ionization cross section on $Z_c$ and $Z_t$.
Neglecting logarithmic contributions,  both the Coulomb 
and the transverse cross sections are proportional to $(Z_t/Z_c)^2$.

\subsection{Beam-gas collisions}

The beam-gas lifetime can be expressed in terms of the ionization
cross section $\sigma _i$, and the density $\rho _i$
of the molecules of type $i$ remaining in the beam pipe:

\begin{eqnarray}
\tau ^{-1}=  \Sigma_i~ (\sigma _i \times \rho _i \times c). 
\end{eqnarray}

The calculations presented below
were made for the K-shell ionization.
This is fully justified as, even for  the bunch temperatures, where the 
Boltzmann proportion of atoms in various excited states 
has to be  taken into account, the lifetime of
these excited states is sufficiently short for 
this  approximation to be  valid - independently  of 
the temperature stabilization efficiency.
%\newpage
%width=10cm,bbllx=-29988 ,bblly=274 , bburx=1012 ,bbury=30780}
%bbllx=-29988 ,bblly=274 , bburx=1012 ,bbury=30780}
\begin{figure}[ht]
\begin{center}
\leavevmode
\epsfig{file=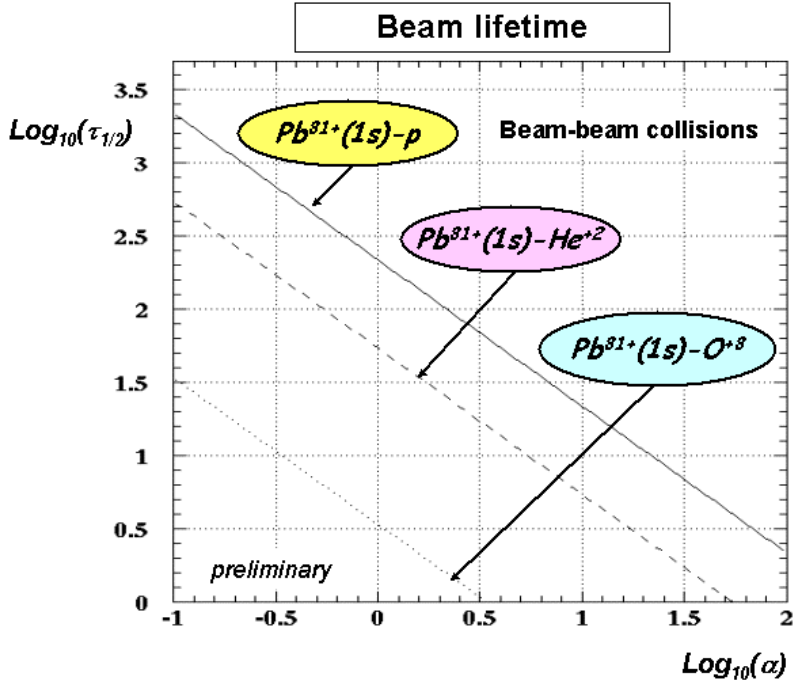,angle=0.,width=13.0cm}
%\vspace{1.0cm}
\end{center}
\caption{The collision lifetime of the $Pb^{81+}$ ions as a function of the 
electron-proton  collison luminosity.
        For the $Pb^{80+}$ ions the luminosity increases  and the lifetime decreases 
by a factor of 2. }
\label{luminosity}
\end{figure}

The calculated ionization cross sections 
for the residual gas molecules listed e.g in 
\cite{Rossi} are large. 
They span the range of $30-2000$ barns 
depending on the type of the molecules 
(the lowest cross section corresponds to the $H_2$ and the highest to the 
CO$_2$ molecules). The beam-loss cross sections for molecules 
containing high $Z$ atoms are significantly higher 
than those
for fully-stripped ions. Therefore,  the vacuum requirements for partially 
stripped ions are  the most demanding ones.
The maximum allowed   
densities  of gas molecules to keep alive
the beam of $Pb^{80+}$ ions 
over $\tau = 10$ hours were found to be 
7 to 200 times lower  than the 
corresponding maximum allowed concentration
for storing fully-stripped lead ions \cite{Jowett}.
The span of these factors reflects the underlying  feature of charge 
coherent processes, which make partially-stripped ion beams singular - 
the strong $Z_t^2$ dependence of the ionization  cross sections.
Therefore, in order to achieve long stores of  partially-stripped ions, 
it is indispensable
to reduce, as much as possible, the contribution of
gasses containing high Z atoms, in particular the 
density of the CO and the CO$_2$  molecules.

The question if the long stores 
of  $Pb^{80+}$ ions are feasible at the LHC 
will remain open until a sufficient
experience of the LHC operation is acquired.
If the estimation of the concentration of 
molecules in the IPs  of the LHC experiments
by Rossi and Hilleret \cite{Rossi} (scenario after machine conditioning) 
will be confirmed by the real LHC operation conditions, the  
ten-hour-long  stores of partially-stripped  lead ions
will be feasible.
The concentration of the CO$_2$ molecules is critical. Their predicted 
density \cite{Rossi} is only two times smaller 
than the maximum allowed 
density for the  ten-hour-long  store of the $Pb^{80+}$ beam, 
leaving only a marginal safety factor.  

Note,  however, that if the vacuum
in the arcs, as it is  expected, will be better
than that in  the interaction regions, for 
which the predictions \cite{Rossi} have been made, 
the beam lifetime will increase.

\subsection{Beam-Beam collisions of partially-stripped ions}

From the above discussion, it is evident that only the high-$Z$
carriers of the electrons can be stored at LHC for 
sufficiently long time. 
The question, which is addressed in this section is: what is the 
luminosity lifetime if the beams of partially-stripped
ions are colliding with each other? 

At first symmetric 
collisions are considered (i.e., those of the same
ions  circulating in the LHC rings in the opposite 
directions).
 The evolution of the number of ions can be calculated 
using the following equation:
\begin{eqnarray}
-1/N \times dN/dt = n_{IR} \times L \times \sigma _{ion} /(k \times N) = \lambda,
\end{eqnarray} 
where $N$ is the number of ions per bunch, 
$k$ is the number of bunches, $\sigma _{ion}$ is the ionization 
cross section and $L$ is the luminosity.

The luminosity evolves with time according to:
\begin{eqnarray}
L(t) = L(0)/(1+ \lambda t)^2.
\end{eqnarray}

Using the Anholt and Becker theory for $\sigma _{ion}$
and assuming the standard
machine parameters: $N = 9 \times 10^7$, $L = 10^{27} cm^{-2}s^{-1}$, 
$k=608$, and
$n_{IR}=2$, the calculated luminosity lifetime  
is found to be smaller than  one  minute. Thus, colliding of beams 
with the same Z is excluded.

\subsection{Allowed collision schemes}

The results presented in the previous section demonstrate clearly, that even
if an effective  beam of electrons carried by high-Z ions
can be stored in  the LHC, 
such a beam cannot be 
used for the high-luminosity collisions of electrons with high-Z nuclei.
Indeed,  the ionization cross section driving the beam losses is huge. 
It  reaches a fraction of Mb for the highest-$Z$ ions
at the maximum LHC energy. This result does not, however,  exclude
other possible collision schemes. 

Indeed, 
the drop of the ionization cross section for Z-asymmetric collisions, 
which is governed  by the $(Z_t/Z_c)^2$ dependence is impressive.
For example, the ionization 
cross section of $Pb^{80+}$  ions colliding 
with the lead ion beam is 
almost 10$^4$ higher than that for collisions with protons.
The asymmetric collisions in which 
two beams of unequal magnetic rigidity are colliding 
are thus allowed - the only 
consequence is  that the effective energies  of ions
circulating in the opposite
beams will be unequal.
The usefulness  of the beam of partially-stripped ions
is thus restricted   to the highly asymmetric
collisions in which the electrons, attached to the highest-Z ions, 
collide with protons or fully-stripped low-Z ions.

In Fig. \ref{luminosity}, the logarithm of the luminosity lifetime 
is plotted against the logarithm of the luminosity 
for collisions of $Pb^{81+}$ ions with protons and low-$Z$, fully 
stripped ions. The luminosity is expressed 
units of $10^{27}$ cm$^{-2}$s$^{-1}$ (the Pb-Pb collision units).
For example, 
if one requires the lifetime to be ten  hours, 
then the luminosity
which can be achieved in the  $Pb^{81+}- proton$ collisions 
will reach the value of $0.4 \times 10 ^{29}$ cm$^{-2}$s$^{-1}$.
Collisions with light ions, with similar luminosity
lifetime,  will also be possible,  albeit
at the price of reduced luminosity. This reduction will be partially 
compensated by the A-times increase of the cross section for 
point-like processes (not taken into account in Fig. \ref{luminosity}).
The rate of point-like collisions of electrons with nuclei scales 
as $A/Z^2$. For example,  the rate of 
the electron-oxygen point-like collisions 
will be reduced by a factor 4 with respect to the electron-proton collision-rate
at the same beam life-time limit.

It is worthwhile to note,  that the integrated  electron-proton luminosity  
used in the first measurements of the structure
functions at small $x$ at HERA \cite{F2} can be 
be collected in  two,   ten-hour-long 
$Pb^{80+}-p$ collision runs at LHC.

\section{The PIE@LHC collider}

\subsection{The running mode}

The physics programme of the electron-proton and
the electron-nucleus collisions, 
must be introduced in a non-invasive way. The merits of this proposal  
must be demonstrated while avoiding any interference with the 
canonical running mode of the LHC collider and while avoiding 
any interference with  optimal tuning of the LHC detectors 
for the standard LHC hadronic-beam programme.

Within the canonical LHC running scenario, 
it is very likely,  that the first Pb-Pb collision runs will be 
followed by the proton-Pb runs.  If, at this stage, the machine
vacuum is controlled to the level discussed in section 5.2 , 
the beam of fully-stripped lead ions could  be replaced by the beam
of partially-stripped
lead ions. 
Collisions of electrons could be recorded in such runs together with 
$proton-Pb$ ($ion-Pb$) collisions.

Those of the LHC
experiments, which would  not be interested in the PIE@LHC collider
programme,  could ignore the parasitic electron-proton collisions altogether.
The only noticeable difference for them 
will be a residual  reduction
of the CM-energy of the $proton-Pb$ ($ion-Pb$) collisions by
an $r=\sqrt{80/82}$ factor. 
Since the rate     
of $proton-electron$ ($ion-electron$) collisions 
will be significantly lower than that of the 
$proton-Pb$ collisions,
their triggering, filtering  
and registering would  consume only an insignificant  
fraction of the data collection resources,
which may remain  
devoted to recording  $proton-Pb$ collision events. 

The only required action of 
those of the LHC experiments 
which will be willing to analyze the electron-proton collision events 
is to include  a dedicated trigger  capable to select a ``topologically
spectacular'' electron-proton collision. The
trigger and the data-acquisition model discussed in \cite{krasny}
anticipates already a clash-less  absorption  of the PIE@LHC collider programme in 
a way which is transparent to the mainstream physics programs
of the present LHC detectors.  

\subsection{The collider parameters}

The parameters of the  PIE@LHC collider 
for electron-proton collisions are summarized below:
\begin{itemize}
\item
CM-energy - $200$ GeV,
\item
luminosity - $0.4 \times 10^{29}$ cm$^{-2}$ s$^{-1}$,
\item
number of bunches - $608$,
\item
number of ions per bunch - $9.4 \times 10^7$,
\item
number of protons per bunch - $4 \times 10^9$,
\item
transverse emittance - $ \leq 1.5$ $\mu m$, 
\item
$\beta ^{*} \geq 0.5~m$ (depending on the reduction factor
of the beam emittance).
\end{itemize}

\subsection{The kinematic domain} 

\begin{figure}[ht]
\begin{center}
\leavevmode
\epsfig{file=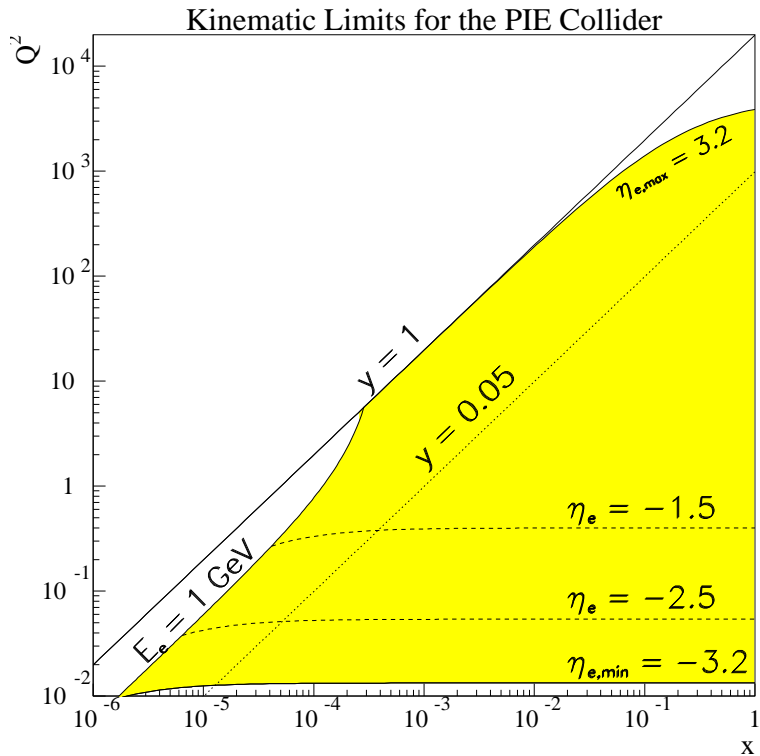,angle=0.,width=11.0cm}
%\vspace{1.0cm}
\end{center}
\caption{Kinematic domain of the PIE collider - the pseudorapidity  acceptance
for  the  electron-proton collisions.}
\label{Janusz1}
\end{figure}

\begin{figure}[ht]
\begin{center}
\leavevmode
\epsfig{file=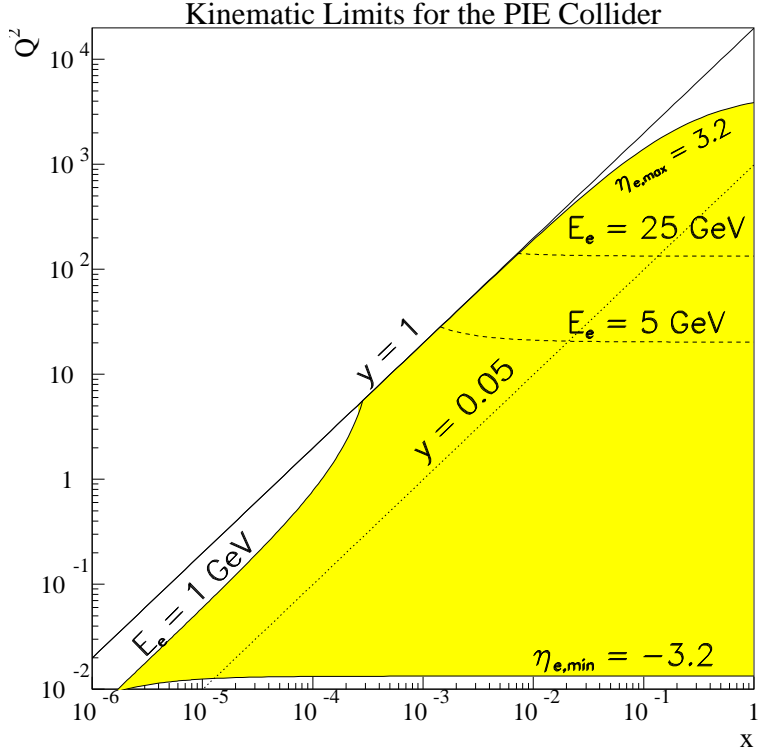,angle=0.,width=11.0cm}
%\vspace{1.0cm}
\end{center}
\caption{Kinematic domain of the PIE collider - the scattered-electron energies
for the  electron-proton collisions.}
\label{Janusz2}
\end{figure}

The kinematic ($x,Q^2$) domain which is accessible for the PIE 
collider is shown in Fig. \ref{Janusz1} and in Fig. \ref{Janusz2}.
 %\newpage
%width=10cm,bbllx=-29988 ,bblly=274 , bburx=1012 ,bbury=30780}
%bbllx=-29988 ,bblly=274 , bburx=1012 ,bbury=30780}

The covered region extends over 5 orders of magnitude
in $x_{Bj}$ and $Q^2$  and over 3 orders of magnitude
in the perturbative-QCD region. The measurement domain is 
limited by the pseudorapidity
$\eta= \pm 3.2$ lines reflecting the geometrical acceptance limit
of measuring the scattered electron in the ATLAS detector \cite{ATLAS},  
and by the $E_e=1$ GeV  line reflecting the
minimum energy 
at which the  electrons can be comfortably identified and measured.

In order to illustrate the  achievable resolution 
in $x_{Bj}$ the $y~=~0.05$ line
is drawn. Below this line, the  $x_{Bj}$ resolution is larger than 100 $\%$. 
In the ($x,Q^2$) domain 
above the   $\eta~=-1.5$ line most of scattered electrons will 
be detected in the  high precision  barrel detector.
It must be  stressed,  that over the a  large part of the kinematic 
domain the scattered electron carries a very small
energy.
Even if, from the point of view of the precision of the measurement 
of the event kinematics,  this is can be handled easily by reconstructing 
the track momentum rather than the electron-cluster energy, 
triggering of these events will be  a challenge.
For example - the minimum bias first level trigger
will have to be followed by a dedicated, topological, second level trigger
\cite{krasny}.  
%\newpage
%width=10cm,bbllx=-29988 ,bblly=274 , bburx=1012 ,bbury=30780}
%bbllx=-29988 ,bblly=274 , bburx=1012 ,bbury=30780}

With respect to earlier experiments,
the specialty  of the PIE@LHC  collider,
related to high asymmetry of the energy of the electron and the proton beams,
is that it covers the small $x_{Bj}$ domain for  a broad 
range  of the $Q^2$ values. This feature of the PIE@LHC
 provides an alternative 
to the proposed HERA III programme, which attempts  to cover  
such a domain by extending the measurement of the  scattered electrons  
to very small angles by means of  a new detector \cite{HERAIII}.

\section{Outlook}

The ideas and the calculations presented in this paper
need to be followed by a  dedicated accelerator
expert's analysis.  If no ``showstoppers'' are found, dedicated 
measurements  will have to be made. Even if the decisive  
feasibility proof
of running partially-stripped ion beams  at the LHC have
to be postponed until the LHC collider is operational, 
several key tests  can be done already now at BNL.

The first, and perhaps the most important one, is to  
send the low intensity $Au^{77+}$
beam to the RHIC and to measure  its life time.
This operation requires refraining from  stripping of the last 
two electrons which takes place in the transfer line between AGS and RHIC.

If stable beams are  observed, 
and if the $Au^{79+}-d$ collision runs at the RHIC are scheduled again, then 
it would be highly desirable to 
replace the $Au^{79+}$ by the  $Au^{77+}$ ions. This 
exercise could
determine  the luminosity lifetime for collisions
of partially-stripped ion beam  and would allow a comparison of 
the emittance of the $Au^{77+}$ beam with 
that of the  $Au^{79+}$ beam
at various stages of the beam acceleration in the RHIC complex.

A dedicated
investigation of the proposed beam cooling method could 
involve emittance studies of the AGS $Au^{77+}$ beam  
and studies of its  evolution in the RHIC ring, 
as a function of the acceleration rate. If a reduction of 
the transverse normalized
emittance is  observed, then 
the next step would be to 
try to understand the beam-cooling mechanism  by 
a direct detection of monochromatic photons  
emitted in radiative transitions of 
thermally excited gold ions. The characteristic photon  
energies  of the atomic de-excitations of gold ions  
are expected to be Doppler-shifted by a $\gamma$ factor reflecting 
the acceleration phase of the beam. 
The detector of these photons should
cover the energy range of 0.5-5 MeV and  
the angular range of $0~\leq~\theta~\leq~ \gamma$
collimated along the direction of the beam of partially 
stripped ions.

Observing the  $\gamma$ dependence 
of the rate of the  
monochromatic photons could   
identify the characteristic acceleration phase in which 
the bunch-temperature stabilization takes place. 
Last, but not the least,   
an experimental investigation of the light emission in the process of
the iso-thermal heat evaporation   
could shed some light on the mechanics the light emission 
in supernova explosions.

Even if an effective spontaneous cooling of the $Au^{77+}$ bunches  
is  observed 
such a beam cannot be used in  $Au^{77+}-Au^{77+}$ collisions
because the luminosity lifetime would be too short.
Moreover, recording $ed$ collisions in  $Au^{77+}-d$
runs  would not add much to the
scope  of the BNL physics programme - the equivalent electron beam 
energy is simply too low. Why then bother with such studies at BNL?
The main reason is that if the beam colling method would work, the gain 
for the future eRHIC@BNL  programme would be of high value - 
in particular if the option of the dedicated QCD-facility 
\cite{eRHIC} could be realized at BNL. In such a facility 
low emittance nuclear beams are indispensable  for using 
nuclei as low-noise, femto-detectors  capable of studying the space-time
evolution of variable partonic configurations at the Fermi distance-scale.

Only if the results of the above tests at BNL would validate
the ideas presented in this paper, it would make sense to 
consider including 
partially-stripped-ion-runs in the CERN LHC programme,
and changing the stripping sequence of ions.

\section{Conclusions} 

The preliminary results presented in this paper suggest  that 
electrons attached to the high-Z ions can be transported
over the LHC acceleration chain and stored at the top LHC energy.

The hybrid beam of monochromatic electrons and their carrier ions
is fragile and can be only used in collisions with protons and fully 
stripped light ions. The luminosity
of electron-proton and electron-light ion collisions,
appears to be sufficient to provide a precise, detector-dependent 
diagnostic of partonic WBBs used in the standard 
LHC physics programme.  

The beam colling method discussed in this paper, if confirmed
experimentally,  may have an important impact on the
acceleration and storage of high energy ion beams and
on the future eRHIC project at BNL.

\section*{Acknowledgment}

I would like to thank Kai Hencken for drawing my attention to ionization
processes as the main limitations for storing partially-stripped ions,
to Leif Ahrens, Steve Peggs and Dejan Trbojevic 
for several valuable suggestions, to Stefan Valkar for 
discussion of  negatively charged hydrogen atoms, 
to Daniel Brandt,  Karl-Heinz Schindl, and John Jowett
for their interest in this  work, to Janusz Chwastowski and Steve
Armstrong for help in preparing the manuscript.


\begin{thebibliography}{9}
  
\bibitem{Rausch} H. Rausch and V. Traubenberg, Naturwiss. {\bf 18}, 417 (1930).

\bibitem{Lanczos} C. Lanczos, Z. Physik {\bf 62}, 518 (1930) and {\bf 68}, 204 (1931). 

\bibitem{Openheimer} J.R. Oppenheimer, Phys. Rev. {\bf 31}, 66 (1928).

\bibitem{Bessonov} E.G. Bessonov, E.V. Tkalya, arXiv:physics/0212100v1, 27 Dec. 2002.

\bibitem{Anholt} R. Anholt and U. Becker, Phys. Rev. {\bf A36}, 4628, (1987).
              
\bibitem{Krause} H.F. Krause et al., Phys. Rev. {\bf A63}, 63, (2001).

\bibitem{Jowett} J.M. Jowett, Ions in the LHC Ring, 
                Proceedings of the LHC project Workshop - Chamonix XIII, CERN-AB-2004-014 ADM.

\bibitem{Rossi} A. Rossi, N. Hilleret, Residual Gas Density Estimation in the LHC
                Experimental Interaction Regions, LHC Project Report 674 (2003).

\bibitem{F2}    I. Abt et al., Nucl.Phys. {\bf B407}, 515 (1993). 


\bibitem{krasny}   M.W. Krasny, A Model 
                       of Dynamic Integration 
                       of ATLAS Detector, Trigger and Software
                       in LHC Data-Taking Environment. {\bf Note II} : A Gauge Model of
                       Data Taking, ATL-COM-GEN-2003-004, CERN, 16 Dec 2003, and references
                       quoted therein. 
                         

\bibitem{eRHIC}    M.W. Krasny, Nucl.Phys.Proc.Suppl. {\bf 105} 185, (2002). 

\bibitem{ATLAS} ATLAS Detector and Physics Performance, Technical Design Report.
                CERN/LHCC/99-15, 25 May 1999.
          
\bibitem{HERAIII} A New Experiment for the HERA Collider. Expression of interest.
               DESY, April 2003.  
              

        
\end{thebibliography}
\end{document}